\documentclass[prd,twocolumn,superscriptaddress,longbibliography,preprintnumbers,floatfix,nofootinbib]{revtex4-1}

\usepackage{url}
\usepackage{xspace}
\usepackage{dsfont}
\usepackage{amssymb}
\usepackage{amsmath}
\usepackage{graphicx}
\usepackage[caption=false]{subfig}
\usepackage[colorlinks=true,citecolor=blue]{hyperref}
\usepackage{multirow}
\usepackage{hhline}

\usepackage{color}
\definecolor{darkred}{rgb}{1.0,0.1,0.1}
\definecolor{darkgreen}{rgb}{0.1,0.7,0.1}
\definecolor{darkblue}{rgb}{0.1,0.1,1.0}

\usepackage{amsmath}

\DeclareRobustCommand{\Sec}[1]{Sec.~\ref{sec:#1}}
\DeclareRobustCommand{\Secs}[2]{Secs.~\ref{sec:#1} and \ref{sec:#2}}

\DeclareRobustCommand{\Fig}[1]{Fig.~\ref{fig:#1}}

\DeclareRobustCommand{\Eq}[1]{Eq.~(\ref{eq:#1})}
\DeclareRobustCommand{\Eqs}[2]{Eqs.~(\ref{eq:#1}) and (\ref{eq:#2})}
\DeclareRobustCommand{\Ref}[1]{Ref.~\cite{#1}}
\DeclareRobustCommand{\Refs}[1]{Refs.~\cite{#1}}

\newcommand{\DCTR}{{\sc Dctr}\xspace}
\newcommand{\OmniFold}{{\sc OmniFold}\xspace}
\newcommand{\Keras}{{\sc Keras}\xspace}
\newcommand{\TensorFlow}{{\sc TensorFlow}\xspace}
\newcommand{\Adam}{{\sc Adam}\xspace}

\begin{document}

\title{A Neural Resampler for Monte Carlo Reweighting with Preserved Uncertainties}

\preprint{MIT-CTP 5224}

\author{Benjamin Nachman}
\email{bpnachman@lbl.gov}
\affiliation{Physics Division, Lawrence Berkeley National Laboratory, Berkeley, CA 94720, USA}

\author{Jesse Thaler}
\email{jthaler@mit.edu}
\affiliation{Center for Theoretical Physics, Massachusetts Institute of Technology, Cambridge, MA 02139, USA}

\begin{abstract}
Monte Carlo event generators are an essential tool for data analysis in collider physics.
To include subleading quantum corrections, these generators often need to produce negative weight events, which leads to statistical dilution of the datasets and downstream computational costs for detector simulation.
Building on the recent proposal of a positive resampler method to rebalance weights within histogram bins, we introduce neural resampling:  an unbinned approach to Monte Carlo reweighting based on neural networks that scales well to high-dimensional and variable-dimensional phase space.
We pay particular attention to preserving the statistical properties of the event sample, such that neural resampling not only maintains the mean value of any observable but also its Monte Carlo uncertainty.
This uncertainty preservation scheme is general and can also be applied to binned (non-neural network) resampling.
To illustrate our neural resampling approach, we present a case study from the Large Hadron Collider of top quark pair production at next-to-leading order matched to a parton shower.
\end{abstract}

\maketitle

\tableofcontents

\section{Introduction}

Data analysis in collider physics relies heavily on simulations to achieve precision.
The state-of-the-art setup includes general purpose event generators like \textsc{Pythia}~\cite{Sjostrand:2014zea}, \textsc{Herwig}~\cite{Bellm:2015jjp}, and \textsc{Sherpa}~\cite{Bothmann:2019yzt}, interfaced with next-to-leading order (NLO) corrections in the strong and/or electroweak couplings.
These higher-order corrections are based on methods such as \textsc{MC@NLO}~\cite{Nason:2006hfa} and \textsc{Powheg}~\cite{Frixione:2002ik} built into the general purpose generators or available as standalone packages like \textsc{MG5\_aMC@NLO}~\cite{Alwall:2014hca} and \textsc{Powheg-Box}~\cite{Alioli:2010xd}.
Outputs from these Monte Carlo generators are then fed into sophisticated detector simulation frameworks, such as those based on \textsc{Geant4}~\cite{Agostinelli:2002hh}.
Event generation and simulation is becoming an increasing relevant computational bottleneck for collider data analysis~\cite{Alves:2017she,Valassi:2020ueh}, particularly for the upcoming High Luminosity Large Hadron Collider (HL-LHC).

Part of the computational burden of Monte Carlo event production is related to the appearance of negative weight events.
Each collision event produced by the pipeline above has an associated event weight $w_i$.
The weights represent the fact that a single simulated collision does not necessarily correspond to one real event.
With the introduction of NLO corrections from quantum loops, some of these weights can even be negative, and the spread in event weights can get increasingly worse with higher-order calculations.
As discussed more below, these negative weights complicate simulation-based inference and increase the computational demands.

In this paper, we present a \emph{neural resampler} for Monte Carlo reweighting, which removes negative weights while preserving the statistical properties of the event sample.
Our method builds upon the positive resampler approach introduced in \Ref{andersen2020positive}, which uses histograms to determine bin-by-bin reweighting factors.
The new feature of neural resampling is the use of neural networks to determine the reweighting factors, which allows us to work with the full unbinned, high-dimensional (and variable-dimensional) phase space.
Another difference with \Ref{andersen2020positive} is that we pay particular attention to preserving the statistical uncertainties of the event sample, which we accomplish by pairing a local reweighting factor with a local resampling rate.

Let us review why weighted Monte Carlo events pose a computational burden for event generation and simulation.
Here, we use ``generation'' to refer to particle-level event generation (e.g.~\textsc{Pythia}) and ``simulation'' to refer to detector simulation (e.g.~\textsc{Geant4}).
When a particle-level phase space point has a non-trivial spectrum of weights (e.g.\ positive and negative), this implies that a large number of generated events will be needed to achieve statistical accuracy.
The reason is that only the average weight at a given phase space point is physically relevant, so if the local phase space weights have a large variance, then many generated events will be needed to accurately estimate the mean.
In some cases, it might be possible to reduce the occurrence of negative weight events directly in the generation step~\cite{Frederix:2020trv}, though this depends on the specific generator implementation.
Barring that, a large number of particle-level events will need to be passed to the computationally expensive detector simulation step to achieve the desired accuracy.

The idea behind positive resampling~\cite{andersen2020positive} is to remove negative weights using a quasi-local weight rebalancing scheme.
This method does not depend on how the particle-level events are generated and it is relevant anytime there is a non-trivial spectrum of weights (i.e.~even if all the weights are positive but different).
Though this method requires choosing a set of observables for histogram binning, the performance is very good when using a relatively small number of physically motivated observables.
Note that positive resampling does not reduce the number of particle-level events that must be generated in order to achieve the desired statistical accuracy.
Still, by rebalancing the event weights, positive resampling does decrease the number of particle-level events that are subsequently fed through the detector simulation.
This can significantly reduce the computational cost of the full event production pipeline.

With neural resampling, we take advantage of the fact that neural networks, when paired with a suitable training algorithm, are excellent likelihood ratio estimators.
This fact has been exploited in a variety of recent studies in high-energy physics~\cite{Cranmer:2015bka,Brehmer:2018eca,Brehmer:2018kdj,Brehmer:2018hga,Stoye:2018ovl,Andreassen:2019nnm,Brehmer:2019xox,Andreassen:2019cjw,Andreassen:2020nkr,Hollingsworth:2020kjg,Badiali:2020wal}.
The technique presented here is most closely related to the \OmniFold unfolding algorithm~\cite{Andreassen:2019cjw}, which is based on the \DCTR technique for full phase space reweighting~\cite{Andreassen:2019nnm}.
Essentially, one can think of Monte Carlo reweighting as computing the likelihood ratio between the desired distribution and an unweighted baseline.
Neural networks are particularly well suited for this task because they can naturally process the high-dimensional phase spaces encountered in Monte Carlo event generation.
To handle variable-dimensional phase spaces, we take advantage of the particle flow network (PFN) architecture for point cloud learning~\cite{Komiske:2018cqr,NIPS2017_6931}.

After positive resampling or neural resampling, all of the event weights will be non-negative (assuming non-negative cross section), though they may not be uniform.
The positive resampling method has a partial unweighting hyperparameter that determines what fraction of events are given a uniform weight~\cite{andersen2020positive}.
In general, though, (partial) unweighting does not preserve the statistical properties of the event sample.
Specifically, unweighting does preserve the expected mean of any observable, but it does not preserve the sample variance.
As reviewed below, sample variance is the most common way to estimate Monte Carlo uncertainties.%
\footnote{The true variance, which is preserved by unweighting, is typically not accessible.}
Crucially, neural resampling preserves the sample variance via a local resampling rate, without needing to choose any hyperparameters.

After neural resampling, one could optionally use standard unweighting techniques to make all of the weights uniform.  While unweighting decreases the overall statistical power of the generated dataset by removing positive-weight events, it has the benefit of maximizing the per-event statistical power of the surviving subset.  For this reason, there is a computational tradeoff between having weighted and unweighted events, with full unweighting being the best strategy when detector simulation is significantly more costly than event generation.  The presence of negative-weight events, or any non-trivial distribution of weights at a given phase space point, is entirely deleterious and only slows statistical convergence.  For this reason, neural resampling is always beneficial from the computational perspective, whether or not there is a subsequent unweighting step.

As an alternative use of neural networks, Monte Carlo unweighting could be achieved using generative models such as generative adversarial networks~\cite{Goodfellow:2014upx} and variational autoencoders~\cite{2013arXiv1312.6114K,2014arXiv1401.4082J}.
Generative models have been extensively studied to accelerate or augment many aspects of high-energy physics simulations~\cite{deOliveira:2017pjk,Paganini:2017hrr,Paganini:2017dwg,Alonso-Monsalve:2018aqs,Butter:2019eyo,Martinez:2019jlu,Bellagente:2019uyp,Vallecorsa:2019ked,SHiP:2019gcl,Carrazza:2019cnt,Butter:2019cae,Lin:2019htn,DiSipio:2019imz,Hashemi:2019fkn,Chekalina:2018hxi,ATL-SOFT-PUB-2018-001,Zhou:2018ill,Carminati:2018khv,Vallecorsa:2018zco,Datta:2018mwd,Musella:2018rdi,Erdmann:2018kuh,Deja:2019vcv,Derkach:2019qfk,Erbin:2018csv,Erdmann:2018jxd,Oliveira:DLPS2017,deOliveira:2017rwa,Farrell:2019fsm,Hooberman:DLPS2017,Belayneh:2019vyx,buhmann2020getting} and lattice field theory~\cite{Albergo:2019eim,Kanwar:2020xzo,Urban:2018tqv,Nicoli:2020njz}.
These approaches, however, require learning the full phase space density, instead of just the likelihood ratio, which is significantly more complicated than the neural resampling approach presented here.
It is also worth mentioning that neural networks and other machine learning techniques have been studied to improve other components of event generation, including parton density modeling~\cite{Ball:2017nwa,Carrazza:2019mzf}, phase space generation~\cite{Bendavid:2017zhk,Klimek:2018mza,Bothmann:2020ywa,Gao:2020vdv,Gao:2020zvv}, matrix element calculations~\cite{Bishara:2019iwh,Badger:2020uow}, and more~\cite{Monk:2018zsb,Andreassen:2019txo,Andreassen:2018apy}.

The rest of this paper is organized as follows.
In \Sec{stats}, we review the statistics of event weights and introduce a method to preserve statistical uncertainties after local phase space reweighting.  
Then, \Sec{methods} introduces our neural network-based resampling technique and shows how it can remove negative weights while preserving uncertainties.
We demonstrate the performance of neural resampling for two case studies in \Sec{results}:  a simple example of two Gaussians and a realistic collider example of matched NLO top quark pair production.
The paper ends with brief conclusions and outlook in \Sec{conclusions}.

\section{The Statistics of Event Weights}
\label{sec:stats}

\subsection{Review of Weights and Uncertainties}

The output of a generic Monte Carlo event generator is a set of $N$ simulated phase space points $\{x_i\}$ along with their associated weights $\{w_i\}$, sampled from random variables $X$ and $W$, respectively.
One can think of $X\in\mathbb{R}^d$ as representing the event features in a $d$-dimensional phase space.
For simplicity, we work with normalized weights for this discussion:
\begin{equation}
\label{eq:weightnorm}
\sum_{i=1}^N w_i = 1,
\end{equation}
such that $w_i=1/N$ for an unweighted event sample.
It is straightforward to adapt the equations below to alternative weight conventions, and for the case studies in \Sec{results}, we work with samples with initial weights $w_i = \pm 1$.

The expectation value of an observable $\mathcal{O}$ can be estimated through a weighted sum:
\begin{equation}
\label{eq:weightedsum}
\langle \mathcal{O}\rangle \approx\widehat{\mathcal{O}}\equiv \sum_{i=1}^N w_i \, \mathcal{O}(x_i),
\end{equation}
where $\mathcal{O}(x)$ is the value of the observable at phase space point $x$.
For example, $\mathcal{O}(x_i)=\delta{(\text{$x_i$ in bin})}$ for the contents of a histogram bin.
In this discussion, hats always indicate estimates obtained through finite sampling.

The true value $\langle \mathcal{O}\rangle$ can only be obtained in the $N \to \infty$ limit, so the estimate in \Eq{weightedsum} is subject to uncertainties.
The variance of \Eq{weightedsum} is $N$ times the variance of each independent term in the sum:
\begin{equation}
\text{Var}[\widehat{\mathcal{O}}] = N\, \text{Var}[W\mathcal{O}(X)]\,,
\end{equation}
where $W$ is a random variable of which $w_i$ is one realization.
In general, though, the true value of $\text{Var}[W\mathcal{O}(X)]$ is not known \emph{a priori}.
Therefore, we have to estimate the variance through the sample variance:
\begin{equation}
\text{Var}[W\mathcal{O}(X)] \approx \frac{1}{N}\sum_{i=1}^N\Big(w_i\,\mathcal{O}(x_i)\Big)^2 -\bigg(\frac{1}{N}\sum_{i=1}^N w_i\,\mathcal{O}(x_i)\bigg)^2.
\end{equation}
In many cases of interest, the second term is subdominant, such as when $\mathcal{O}(x_i)$ is zero for most phase space points and one otherwise.
Under that assumption, we obtain the standard Monte Carlo uncertainty estimate:
\begin{equation}
\label{eq:MCuncertainty}
\text{Var}[\widehat{\mathcal{O}}] \approx \sum_{i=1}^N \Big(w_i\,\mathcal{O}(x_i)\Big)^2.
\end{equation}

\subsection{Monte Carlo Reweighting}

The idea behind reweighting is to select a subset of phase space points $x_j$ with modified weights $\widetilde{w}_j$ such that \Eq{weightedsum} (and later \Eq{MCuncertainty}) are preserved in the large $N$ limit.
Consider a small patch of phase space such that $\mathcal{O}$ is nearly constant.
In such a region with $N_\text{patch}$ events:
\begin{equation}
\widehat{\mathcal{O}}_\text{patch} \approx  \mathcal{O}(x_{\rm patch}) \sum_{i=1}^{N_\text{patch}} w_i.
\end{equation}
The key observation is that we can replace the $N_\text{patch}$ original events with $N_\text{patch}/K$ events of equal weight $\widetilde{w}_j$,
\begin{equation}
\label{eq:patchweight}
    \widetilde{w}_j = \frac{K}{N_\text{patch}} \sum_{i = 1}^{N_\text{patch}} w_i \approx K\, \langle W\rangle_\text{patch}.
\end{equation}
This replacement preserves the estimate of any observable because:
\begin{align}
\label{eq:patches}
     \sum_{j=1}^{N_\text{patch}/K} \widetilde{w}_j = \sum_{i=1}^{N_\text{patch}} w_i,
\end{align}
where we are assuming $N_\text{patch}/K$ is an integer.
Since $\widehat{\mathcal{O}}_\text{patch}$ is unchanged by this reweighting, the true value of \text{Var}[$\widehat{\mathcal{O}}_\text{patch}$] is also unchanged.
Said another way, one can replace $K$ events in a small patch with a single representative event and achieve approximately the same statistical properties.

As a result, one can reduce the number of events required for expensive detector simulations by picking $K>1$. 
This reduction in statistics is possible for any $K$, though in general one has to be careful when accounting for the uncertainty, which we now discuss.

\subsection{Preserving Uncertainties}
\label{sec:preserve_unc}

The naive uncertainty estimate in \Eq{MCuncertainty} is badly biased if one blindly uses the weights in \Eq{patchweight}.
While it is possible to keep track of the local variance separately from the expected value, this is cumbersome and error prone.
Instead, we advocate choosing $K$ such that
\begin{equation}
\label{eq:w2equal}
\sum_{j=1}^{N_\text{patch}/K} \widetilde{w}_j^2 = \sum_{i=1}^{N_\text{patch}} w_i^2.
\end{equation}
Formally, \Eqs{patches}{w2equal} require $N/K$ to be an integer, but below we will take the continuum limit where this distinction is unnecessary.
With this choice of $K$, one can estimate the variance in the usual way using \Eq{MCuncertainty} without any additional overhead.
This is the key observation that underpins our neural resampling method.

In general, each phase space patch will require a different value of $K$.
To determine this $K_\text{patch}$, it is convenient to rewrite \Eq{w2equal} as:
\begin{align}
\label{eq:rewritew2equal}
\widetilde{w}_j^2=\frac{K_\text{patch}}{N_\text{patch}} \sum_{i=1}^{N_\text{patch}} w_i^2\approx K_\text{patch}\,\langle W^2\rangle_\text{patch}.
\end{align}
Combining \Eqs{rewritew2equal}{patchweight} provides a prescription for choosing $K_\text{patch}$:
\begin{align}
\label{eq:Kdefpatch}
K_\text{patch}\approx\frac{\langle W^2\rangle_\text{patch}}{\langle W\rangle_\text{patch}^2}\,.
\end{align}
Taking the continuum limit, \Eq{Kdefpatch} becomes
\begin{align}
\label{eq:Kdef}
K(X)=\frac{\langle W^2|X\rangle}{\langle W|X\rangle^2}\,,
\end{align}
with expectation values conditioned on the phase space points in $X$.

The above discussion can be encoded in the following practical algorithm to reweight and resample Monte Carlo events while preserving uncertainties:
\begin{enumerate}
\item Estimate $\widehat{W}(X)\approx \langle W|X\rangle$.
\item Estimate $\widehat{W^2}(X)\approx \langle W^2|X\rangle$.
\item Define $\widehat{K}(X)=\widehat{W^2}(X)/\widehat{W}(X)^2$.
\item For each event $i$, keep it with probability $1/\widehat{K}(x_i)$; otherwise discard the event.  Because $\widehat{K}(x_i) \ge 1$ by construction, no event will be repeated.
\item For each kept event, set the new event weight to be $w_i\mapsto \widetilde{W}(x_i)\equiv \widehat{W}(x_i)\, \widehat{K}(x_i)$, which is the continuum limit of $\widetilde{w}$ from \Eq{patchweight}.
\end{enumerate}
The computational benefit of using $\widetilde{W}(x_i)$ over $w_i$ is true even if there are no negative weights.
As with any Monte Carlo method, the accept-reject procedure in step 4 can only preserve \Eqs{patches}{w2equal} in expectation value.
As long as a given phase space point has a non-trivial spectrum of weights, the above reduction will decrease the computational cost of subsequent detector simulation with the same asymptotic statistical properties as captured by the first and second moments.  The procedure above works for any estimation of $\langle W|X\rangle$ and $\langle W^2|X\rangle$, including with histograms.  The next section shows how to estimate these quantities without binning using neural networks.

\section{Neural Resampling}
\label{sec:methods}

As described above, a Monte Carlo generator draws a sample $\{x_i\}$ from $X$.
Each phase space point $x_i$ has an associated weight $w_i$, which can be positive or negative.
Moreover, the weights need not be a function of $X$, meaning that the same phase space point can have different weights, as determined by the Monte Carlo sampling scheme.
The goal of the positive resampler method of \Ref{andersen2020positive} is to rebalance the weights such that each value of $x$ has a unique weight.
Our neural resampler accomplishes this same goal through binary classification with neural networks.

\subsection{Learning Event Weights}
\label{sec:learn_weight}

To learn new event weights, we train a neural network to distinguish between two samples: the original sample $\{x_i\}$ with weights $\{w_i\}$ and a uniformly weighted sample with the same phase space points $\{x_i\}$ but weights set to $1$.
For concreteness, we use the binary cross-entropy loss for this discussion, though other loss functions with the same asymptotic behavior would also work, such as the mean squared error.%
\footnote{\label{footnote:mse}One key difference between binary cross-entropy and mean squared error is that the former cannot learn negative weights.
There are situations, particularly when using fixed-order Monte Carlo generators, where one encounters phase space regions with genuinely negative cross sections.
We performed a preliminary test of this in the context of fixed-order top quark pair production with a parton shower subtraction scheme where, unlike the matched results in \Sec{ttbar}, there are negative phase space regions.
Using the mean squared error loss and linear activation in the final layer, we found good performance in the presence of both positive and negative cross section regions.}

The loss function to be minimized is:
\begin{equation}
\label{eq:loss}
\mathcal{L}[g] = -\sum_{i= 1}^N w_i \log g(x_i) -\sum_{i=1}^N \log\big(1-g(x_i)\big),
\end{equation}
where $g(x)$ is parametrized as a neural network with output range $[0,1]$.
We emphasize that the two sums in this loss function run over the \emph{same} phase space points $x_i$, just with different weights.
This setup is identical to the second step of the \OmniFold unfolding algorithm~\cite{Andreassen:2019cjw}, where a generated dataset is morphed into a weighted version of itself.

Taking a functional derivative of \Eq{loss} with respect to $g(x)$ and setting it equal to zero, one can show that the loss function minimum provides an estimate of $\langle W|X\rangle$:
\begin{equation}
\label{eq:learned_w_given_x}
\frac{g(x)}{1-g(x)} = \widehat{W}(x) \approx \langle W|X\rangle.
\end{equation}
This is just a manifestation of the standard result that asymptotically~(i.e.\ with infinite training data, maximally expressive neural network architecture, and ideal training procedure) the output of a binary classifier approaches a monotonic rescaling of the likelihood ratio; see e.g.~\Refs{Cranmer:2015bka,Brehmer:2018eca,Brehmer:2018kdj,Brehmer:2018hga,Stoye:2018ovl,Andreassen:2019nnm,Brehmer:2019xox,Andreassen:2019cjw,Andreassen:2020nkr,Hollingsworth:2020kjg,Badiali:2020wal}.
In our case, the original sample has asymptotic probability distribution
\begin{equation}
    p_\text{original}(x) = \langle W|x\rangle \, p_\text{uniform}(x),
\end{equation}
where $p_\text{uniform}(x)$ is the phase space prior.
The sample with uniform weights is not a proper probability distribution, since it is not normalized, but corresponds to $N$ times $p_\text{uniform}(x)$.
In this way, we learn local event weights that preserve the estimate of any observable via \Eq{weightedsum}.

\subsection{Learning Uncertainties}

For the case studies in \Sec{results}, we focus on situations where the initial weights are all $\pm c$, for a fixed value of $c$.
In such cases, $\langle W^2|X\rangle=c^2$ by construction and no additional training is needed to perform neural reweighting.

If there is a non-trivial spectrum of weight norms, we can repeat the logic of \Sec{learn_weight} to estimate $\langle W^2|X\rangle$.
This can be achieved using the loss function:
\begin{equation}
\label{eq:loss2}
\mathcal{L}[h] = -\sum_{i=1}^N w_i^2 \log g(x_i) - \sum_{i=1}^N \log\big(1-g(x_i)\big),
\end{equation}
whose asymptotic minimum satisfies:%
\footnote{As a generalization of this result, consider any random variable $X\in\mathbb{R}^d$ and write it as $X=(Y,Z)$ with $Y\in\mathbb{R}^k$ and $Z\in\mathbb{R}^\ell$ with $d=k+\ell$.  Then, one can learn $\langle f(Z)|Y\rangle$ for a fixed function $f$ by treating $Z|Y$ as a set of weights.}
\begin{equation}
\label{eq:learned_w2_given_x}
\frac{h(x)}{1-h(x)} = \widehat{W^2}(x) \approx \langle W^2|X\rangle.
\end{equation}
In this way, we learn local variances that preserve the estimate of observable uncertainties via \Eq{MCuncertainty}.

\subsection{Implementation Details}

To implement neural resampling, the (learned) $\widehat{W}(x)$ and $\widehat{W^2}(x)$ functions from \Eqs{learned_w_given_x}{learned_w2_given_x} are simply inserted into the algorithm described in \Sec{preserve_unc}.
If desired, one could further unweight the samples to make all of the weights uniform; see further discussion in \Ref{andersen2020positive}.
We omit this optional step in our case studies, since full (or partial) unweighting does not preserve the original Monte Carlo uncertainties.

For the following results, neural networks are constructed using rectified linear unit (ReLU) activation functions for hidden layers and the sigmoid function for the last layer to be the output in $[0,1]$.
All models are implemented in \Keras~\cite{keras} with the \TensorFlow backend~\cite{tensorflow} and trained using the cross-entropy loss with the \Adam~\cite{adam} optimizer.
For the one-dimensional example, the neural network is fully connected with three hidden layers, 128 nodes per layer, and trained for 10 epochs.
The higher-dimensional example uses PFNs~\cite{Komiske:2018cqr}, based on the deep sets architecture~\cite{NIPS2017_6931}, using the default parameters from \url{https://energyflow.network/} and trained for 100 epochs.
None of the hyperparameters have been optimized for these studies.

\section{Case Studies}
\label{sec:results}

To illustrate the potential of neural resampling, we present two case studies.
First, we consider an example involving two Gaussians that highlights the potential of our method in a simple case where the optimal results can be obtained analytically.
Then, we consider a realistic collider example of top quark pair production ($t\bar{t}$) at NLO matched to a parton shower, which allows us to demonstrate the robustness of our method to multi-dimensional (and variable-dimensional) phase space. 

For each of the following examples, we work with initial event weights of $w_i = \pm 1$, since this is the typical output of unweighted Monte Carlo generators.
To map onto the discussion in \Secs{stats}{methods}, one would need to normalize these weights to satisfy \Eq{weightnorm}.
Note that $\langle W^2|X\rangle = 1$ by construction, so we can skip the neural network training step in \Eq{loss2}.

\subsection{Two Gaussians}
\label{sec:gauss}

Our first case study involves two Gaussian distributions, one with positive weight and one with negative weight.
Let $X_0\sim\mathcal{N}(0,1)$ and $X_1\sim\mathcal{N}(0,0.5)$, where $\mathcal{N}(\mu,\sigma)$ represents a Gaussian distribution with mean $\mu$ and standard deviation $\sigma$.
The weight for $X_0$ events is $+1$ while the weight for $X_1$ events is $-1$.
We consider the case where the $X_0$ events happen $3$ times more often than the $X_1$ events.
The following results are based on 4M samples from $X$, i.e.\ 3M positive weight events from $X_0$ and 1M negative weight events from $X_1$.

\begin{figure*}[t]
\centering
\subfloat[]{
\label{fig:gauss_dist}
\includegraphics[scale=0.4]{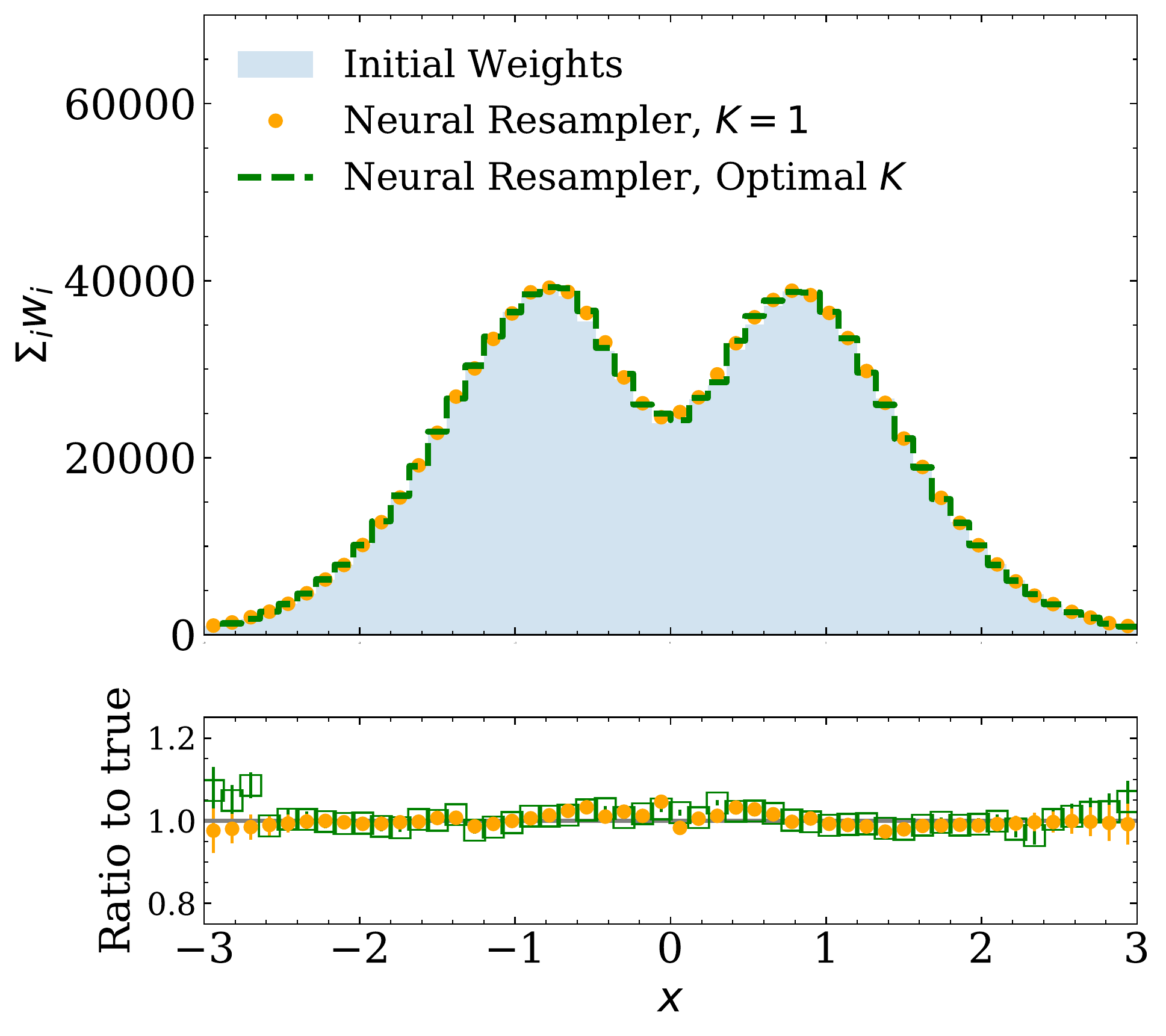}
}
$\qquad$
\subfloat[]{
\label{fig:gauss_weight}
\includegraphics[scale=0.4]{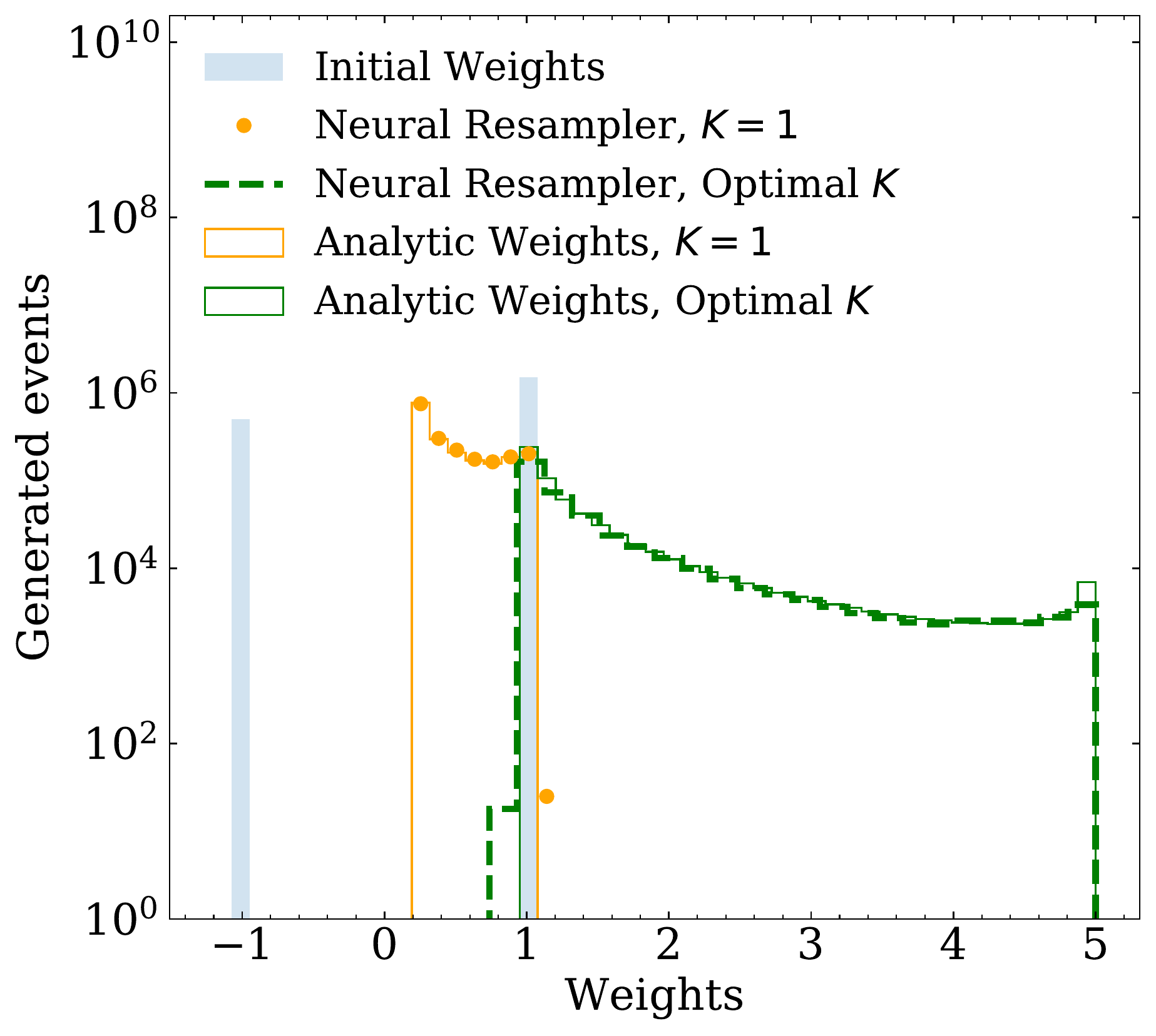}
}
\\
\subfloat[]{
\label{fig:gauss_error}
\includegraphics[scale=0.4]{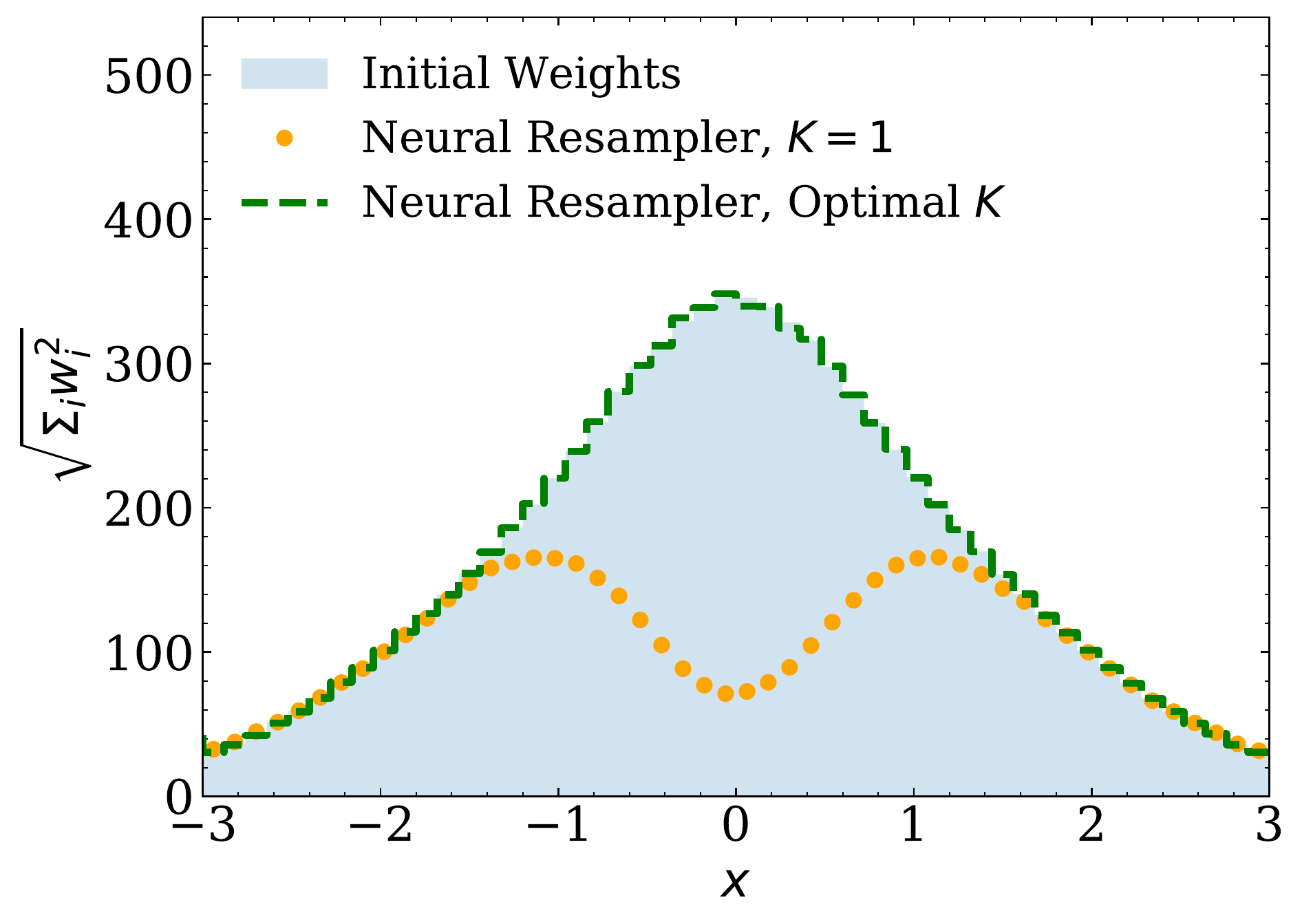}
}
$\qquad$
\subfloat[]{
\label{fig:gauss_events}
\includegraphics[scale=0.4]{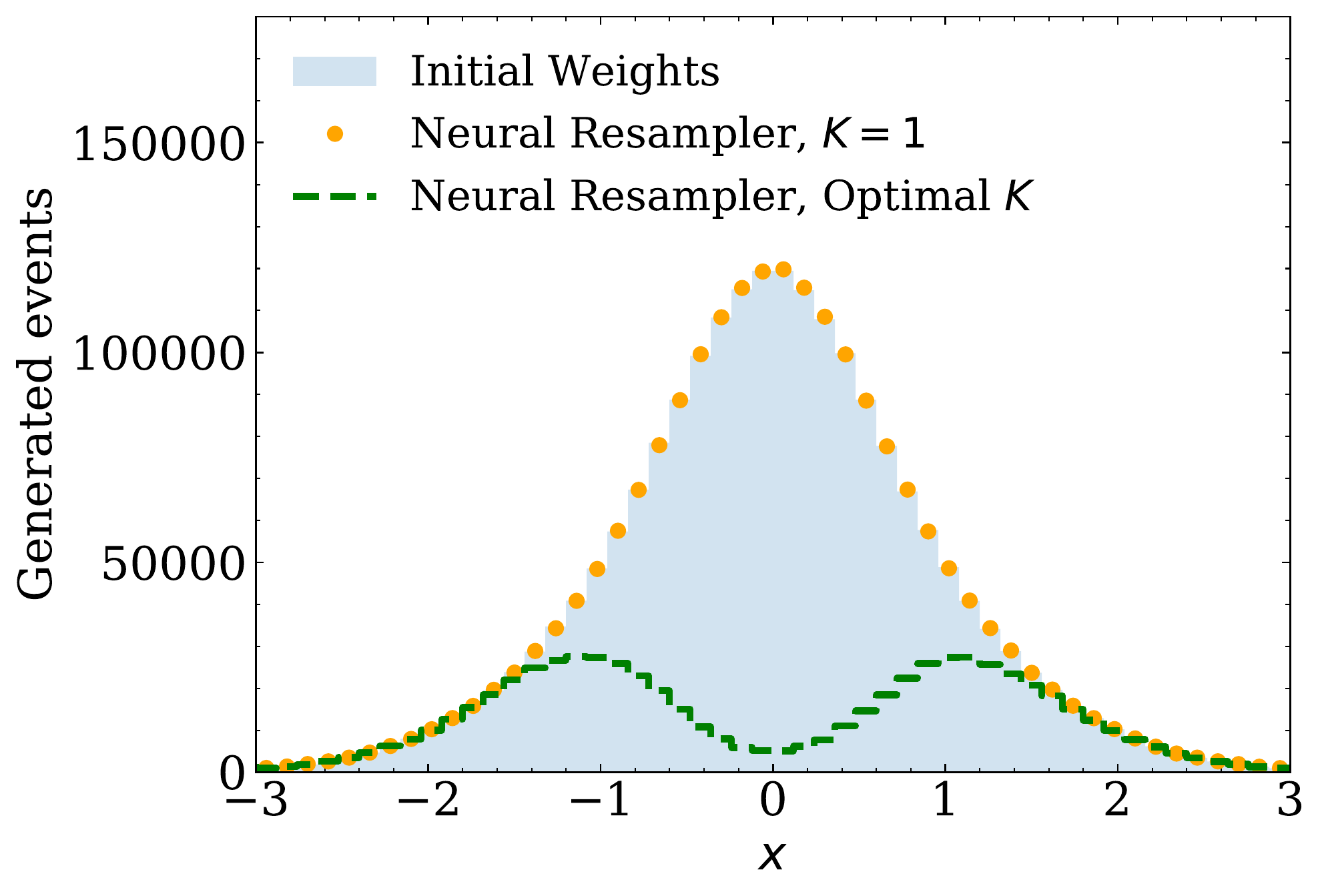}
}
\caption{
Demonstration of neural resampling in a simple one-dimensional example of two Gaussians.
(a) The (rescaled) cross section for observable $x$, estimated by summing the event weights in each histogram bin.
(b) The distribution of the event weights, including solid lines corresponding to the analytic expectations.
(c)  The (rescaled) uncertainties for observable $x$, estimated by summing the squared event weights in each histogram bin and taking the square root.
(d)  The number of events as a function of $x$.
The curves correspond to (solid blue) the original event sample with positive and negative weights, (orange dotted) neural resampling with $K = 1$ such that the cross section is preserved, and (green dashed) neural resampling with the optimal $K$ value in \Eq{Kdef} such that the cross sections and uncertainties are preserved.  Note that the ratio between the orange and green curves in \Fig{gauss_events} is $K(x)$, which is the local factor describing how many fewer events are needed after resampling.
}
\label{fig:gauss}
\end{figure*}

The weighted histogram of $X$ is shown by the blue solid distribution in \Fig{gauss_dist}, corresponding to a Gaussian with a dip in the middle.
After neural resampling with $K = 1$, shown by the orange dotted curve, we obtain the same true distribution up to statistical fluctuations.
As desired, subsampling with the optimal value of $K$ from \Eq{Kdef} does not change the probability density, as shown by the green dashed curve.
This subsampling does change the uncertainties, though, as discussed more below.
We therefore conclude that neural resampling has successfully preserved the cross section via \Eq{weightedsum}.

A nice feature of this example is that the correct unbinned weights are computable analytically using the asymptotic formulas in \Sec{stats}.
In \Fig{gauss_weight}, we show the original weight distribution together with the positive neural resampling weights using both $K = 1$ and the optimal value of $K$.
While the original weights are both positive and negative, neural resampling yields strictly positive weights.
We see that the finite sampling matches the expected analytic weight distributions, which is a non-trivial cross check of our neural reweighting code.
Note that even though a binning and finite sampling are chosen to represent the data in \Fig{gauss}, all of these distributions are fundamentally unbinned.

In \Fig{gauss_error}, we show the distribution of uncertainties using the standard Monte Carlo estimate based on summing the squared weights.
With $K=1$, the uncertainties are substantially underestimated relative to the original distribution, especially in the vicinity of $x = 0$, where there are large cancellations between negative and positive weights.
Subsampling with the optimal $K$ restores the original uncertainties, as desired.
We therefore conclude that neural resampling has successfully preserved the uncertainties via \Eq{MCuncertainty}.

As shown in \Fig{gauss_events}, subsampling yields a significant savings in terms of the number of events required to obtain the same statistical properties as the original sample.
Only one third of the events are needed to capture the same behavior of the original events, with the savings greatest near $x = 0$ where there are larger relative uncertainties.

This two Gaussian example highlights the efficacy of neural resampling in a simple one-dimensional example.
We now turn to a case of relevance to collider physics where the phase space is multi-dimensional.

\subsection{Top Quark Pair Production}
\label{sec:ttbar}

\begin{figure*}[t]
\centering
\subfloat[]{
\label{fig:matched_isr_dist}
\includegraphics[scale=0.4]{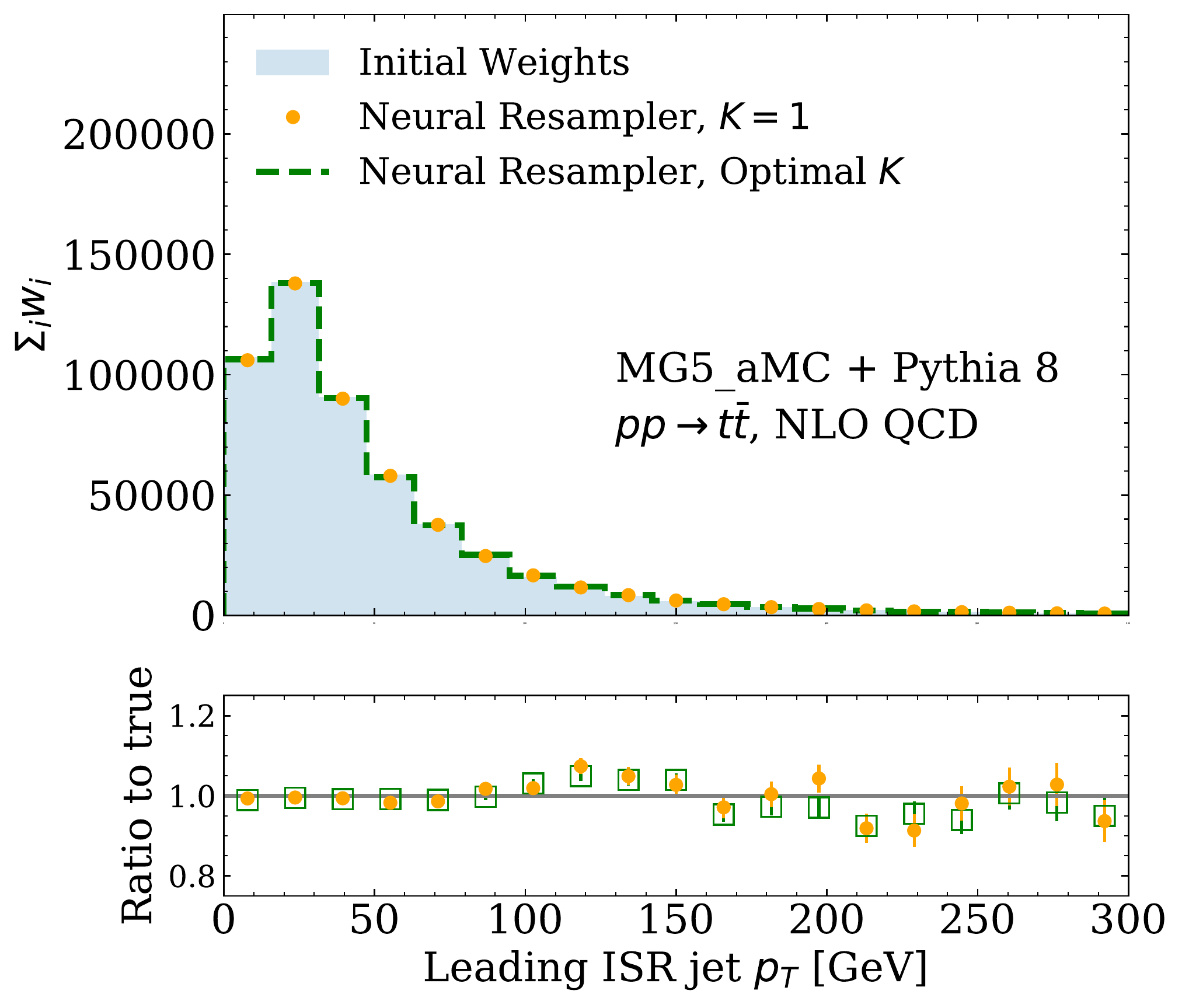}
}
$\qquad$
\subfloat[]{
\label{fig:matched_isr_weight}
\includegraphics[scale=0.4]{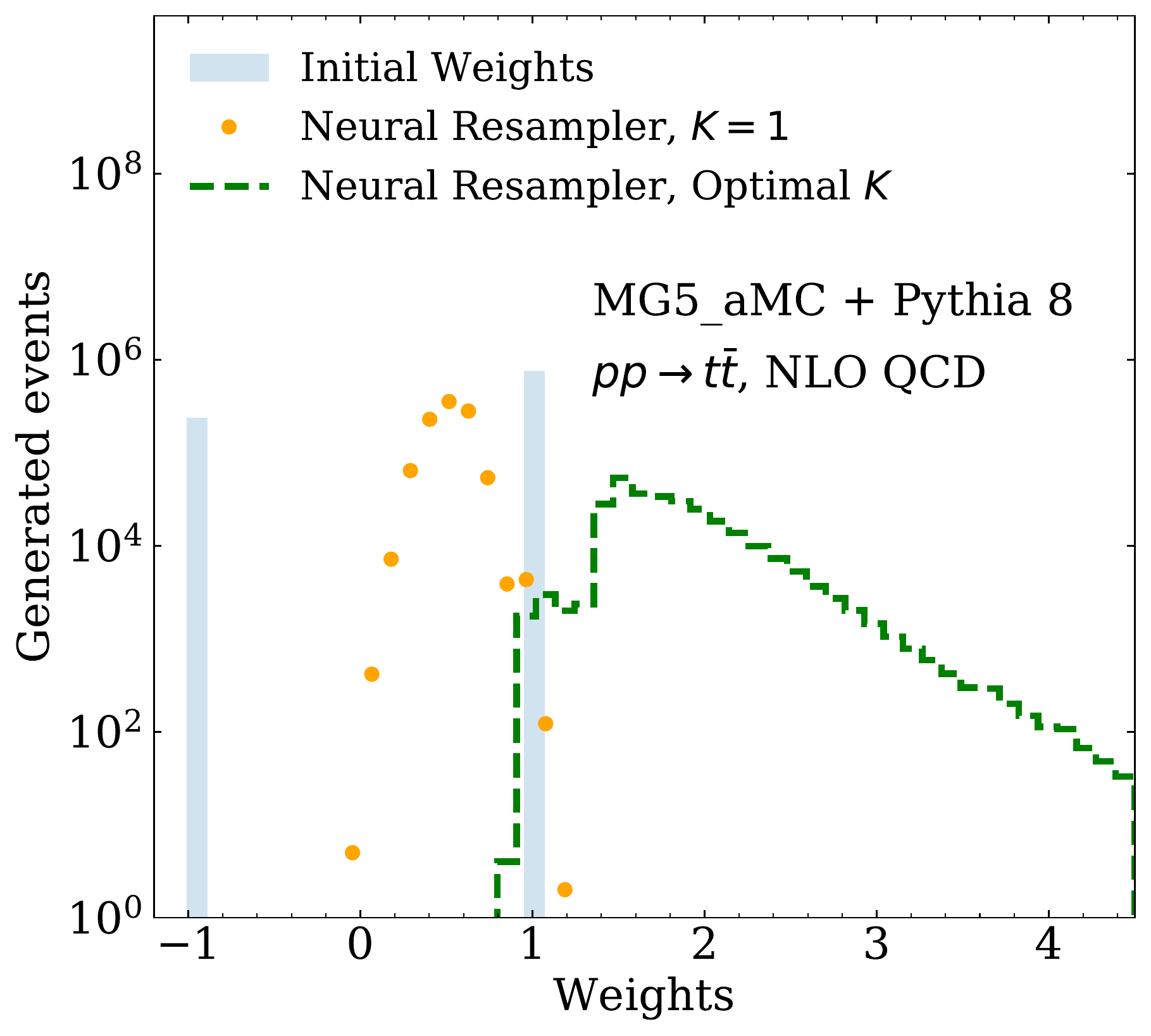}
}
\\
\subfloat[]{
\label{fig:matched_isr_error}
\includegraphics[scale=0.4]{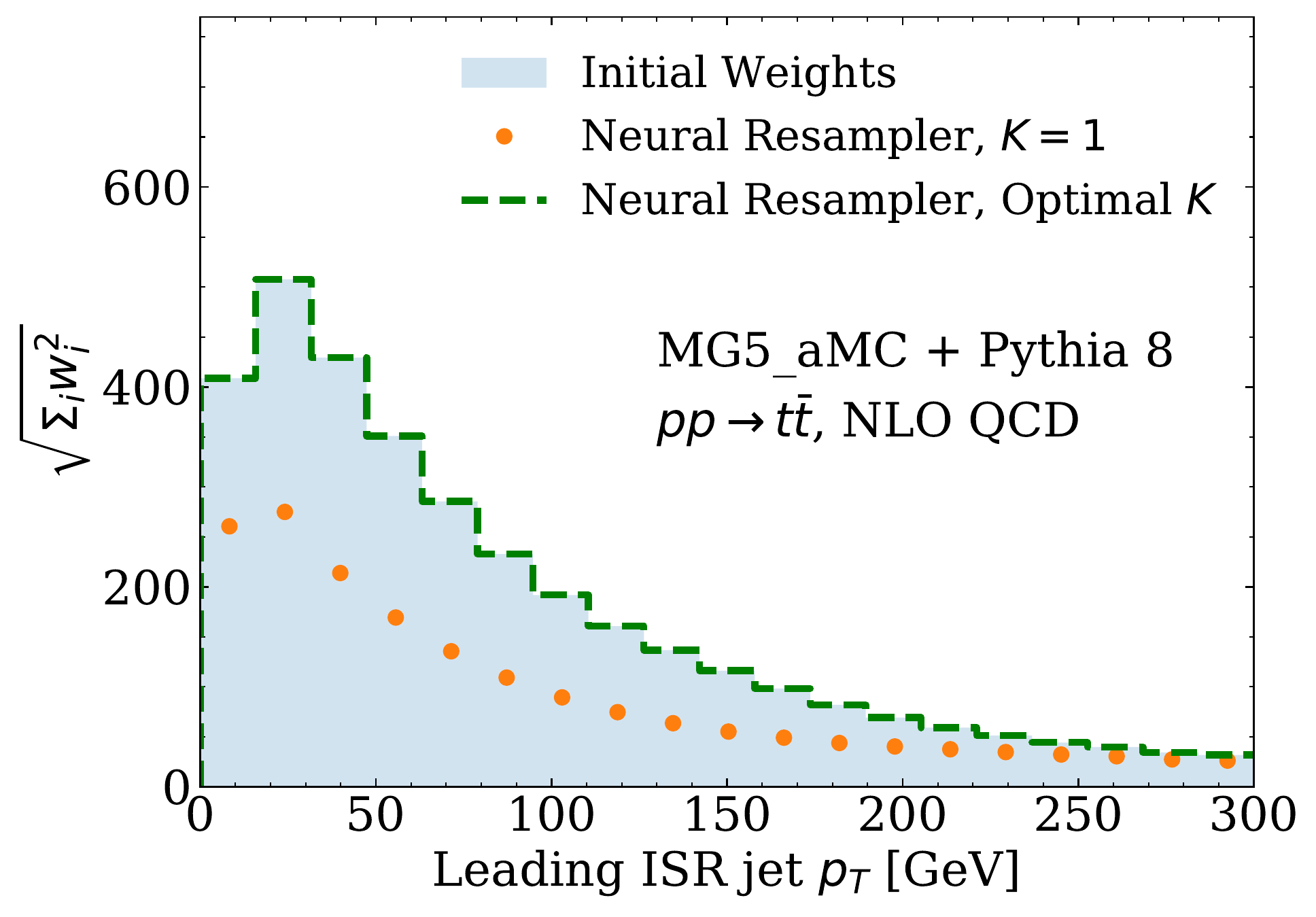}
}
$\qquad$
\subfloat[]{
\label{fig:matched_isr_events}
\includegraphics[scale=0.4]{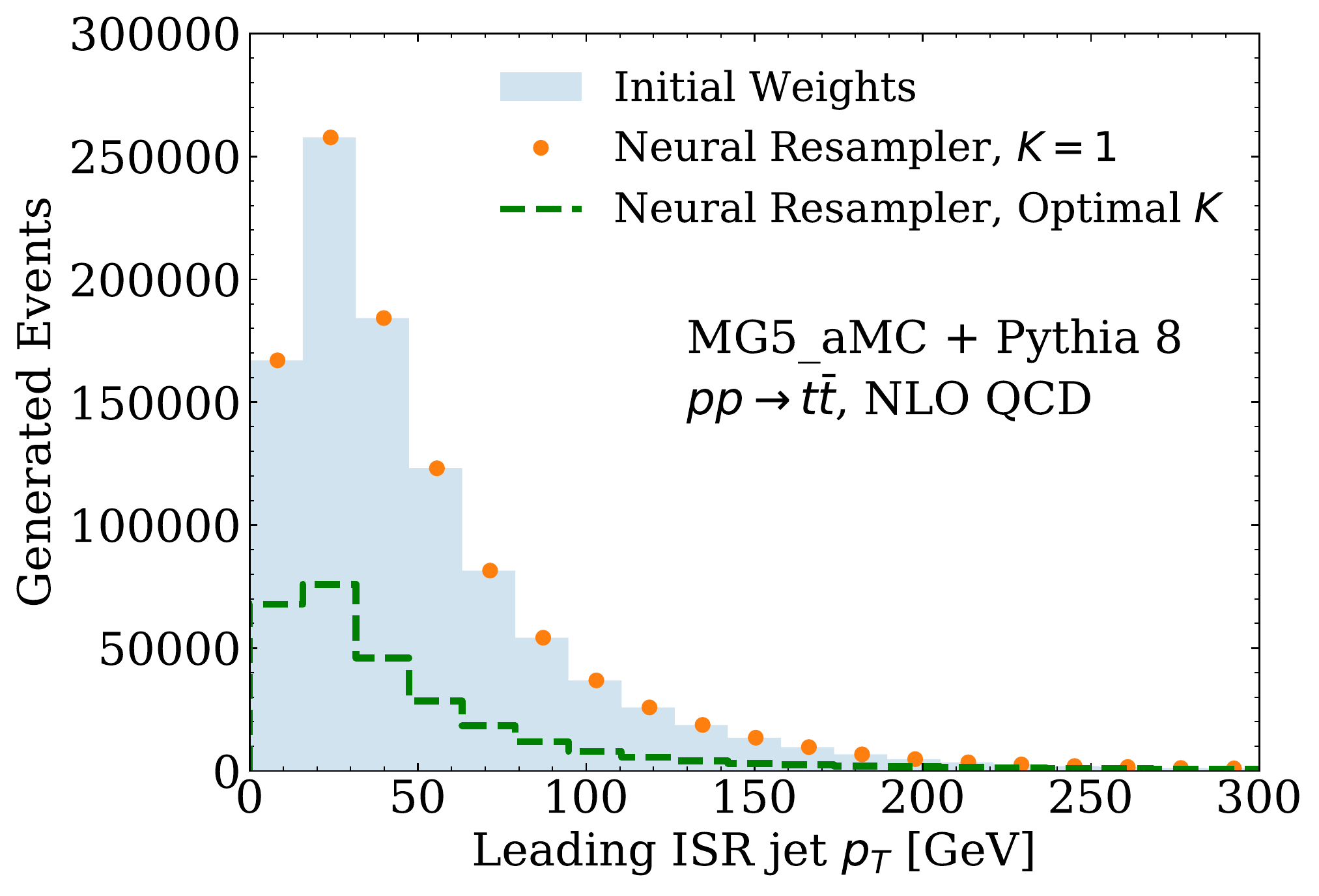}
}
\caption{
Demonstration of neural resampling in a realistic collider example of top quark pair production at NLO matched to a parton shower.
(a) The (rescaled) cross section for the leading ISR jet $p_T$.
(b) The distribution of the event weights.  The optimal $K$ spectrum steeply falls out to about 10, but the distribution is truncated at 4.5 to aid the comparison with the other histograms.
(c)  The (rescaled) uncertainties for $p_T$.
(d)  The number of events as a function of $p_T$.
As in \Fig{gauss}, the curves correspond to (solid blue) the original event sample with positive and negative weights, (orange dotted) neural resampling with $K = 1$, and (green dashed) neural resampling with the optimal $K$ value in \Eq{Kdef}.}
\label{fig:matched_isr}
\end{figure*}

\begin{figure*}[t]
\centering
\subfloat[]{
\label{fig:matched_njet_dist}
\includegraphics[scale=0.4]{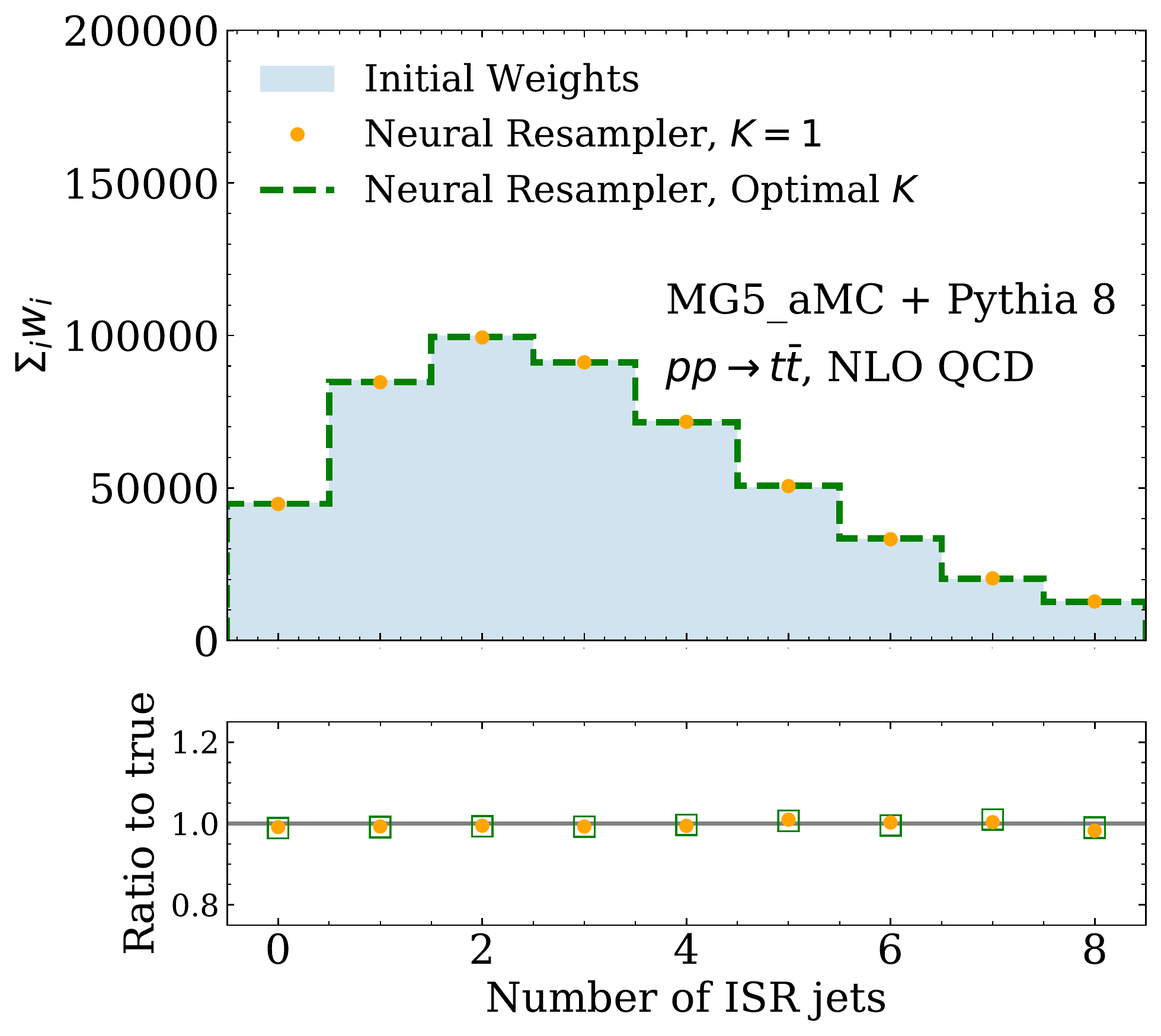}
}
$\qquad$
\subfloat[]{
\label{fig:matched_njet_weight}
\includegraphics[scale=0.4]{wisrpt_full.pdf}
}
\\
\subfloat[]{
\label{fig:matched_njet_error}
\includegraphics[scale=0.4]{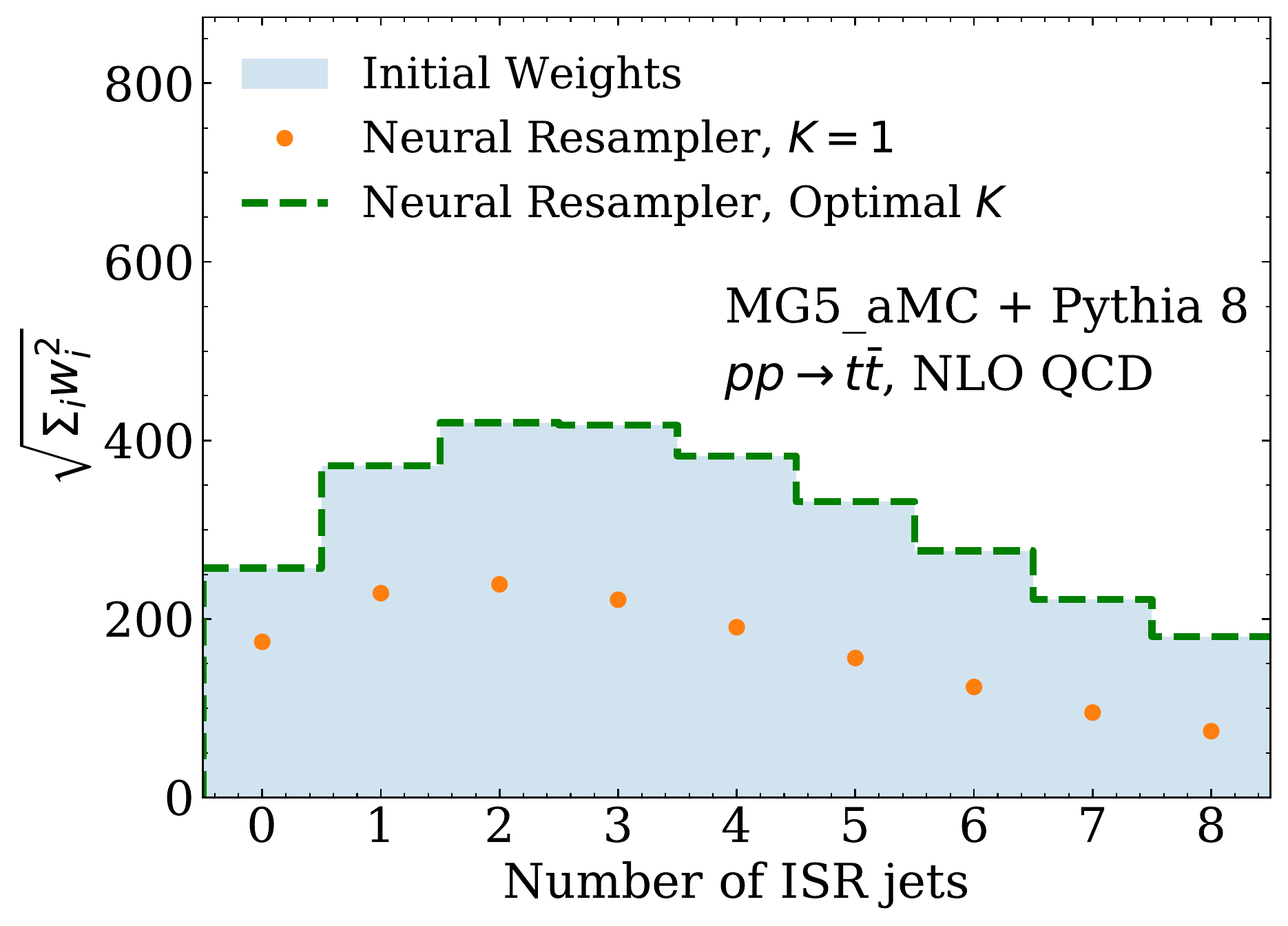}
}
$\qquad$
\subfloat[]{
\label{fig:matched_njet_events}
\includegraphics[scale=0.4]{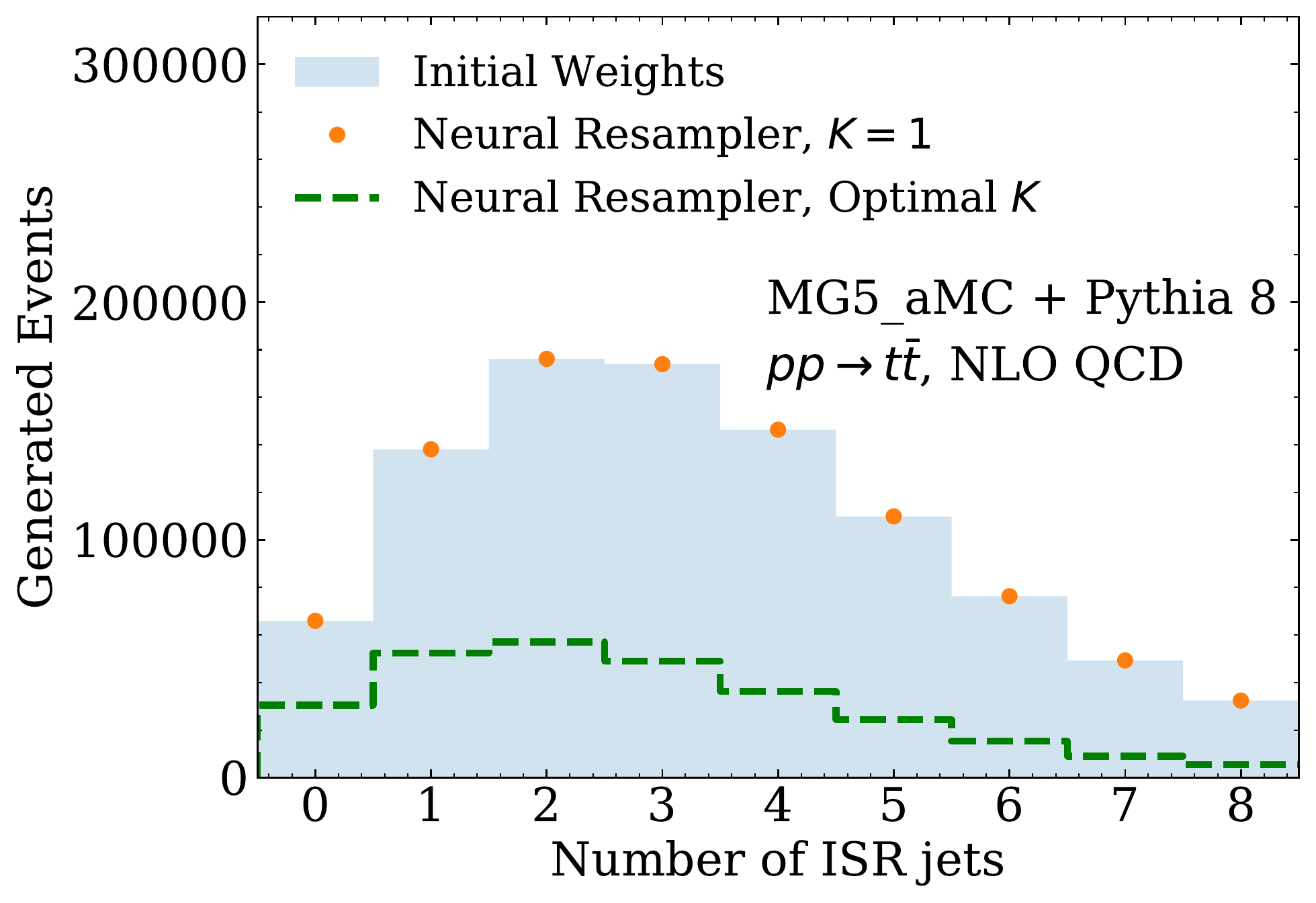}
}
\caption{The same as \Fig{matched_isr}, but now plotting the number of ISR jets with $p_T > 10$~GeV.
Note that the weight distribution in (b) is identical to \Fig{matched_isr_weight}, since neural resampling acts on the full unbinned phase space.
}
\label{fig:matched_njet}
\end{figure*}

Our realistic collider case study is based on top quark pair production at NLO in quantum chromodynamics.
At fixed order in an expansion in the strong coupling constant $\alpha_s$, it is well known that cross sections can become negative due to the breakdown of perturbation theory in the vicinity of soft/collinear singularities (see footnote \ref{footnote:mse}).  
These unphysical phase space regions can be regulated by matching to a parton shower, rendering the differential cross section to be positive.
In addition, matching improves the accuracy of the cross section prediction by including resummation effects beyond NLO.

The analysis below is based on 2M total events generated as follows.
Fixed-order $t\bar{t}$ production is generated using MG5\_aMC@NLO 5.2.7.2~\cite{Alwall:2014hca} interfaced with the NNPDF 2.3 NLO parton density function set~\cite{Ball:2012cx}.
This fixed-order calculation uses \textsc{MadLoop}~2.7.2~\cite{Hirschi:2011pa,Alwall:2014hca}, \textsc{Ninja}~1.2.0~\cite{Mastrolia:2012bu,Peraro:2014cba}, \textsc{CutTools}~1.9.3~\cite{Ossola:2007ax}, and \textsc{OneLoop}~3.6 ~\cite{vanHameren:2010cp,vanHameren:2009dr}.
In these samples, the top quarks are forced to decay leptonically via $t\bar{t}\rightarrow b\bar{b}\mu^+\mu^-\nu\bar{\nu}$.
Using the default FKS subtraction scheme~\cite{Frixione:1995ms,Frixione:1997np}, the resulting events are matched with the \textsc{Pythia} 8.230~\cite{Sjostrand:2014zea,Sjostrand:2006za} parton shower, keeping the default shower settings.
Initial state radiation (ISR) is generated both from the NLO calculation and from the subsequent parton shower matching.

The resulting events in the \textsc{HepMC}~2.06.09~\cite{Dobbs:684090} format are processed with the \textsc{HepMC} reader module of \textsc{Delphes}~3.4.2~\cite{deFavereau:2013fsa}.
This setup is also used to cluster $R = 0.4$ anti-$k_t$ jets~\cite{Cacciari:2011ma} with \textsc{FastJet}~\cite{Cacciari:2005hq}.
Jets originating from a $b$-quark are identified as such using the flavor tagging module in \textsc{Delphes}.
Jets are labeled as ``ISR'' if they are not $b$-jets.
We suppress the overall cross section information and take the weights to be $\pm 1$.
After neural resampling, it is straightforward to rescale all of the weights to have the proper dimensions of a cross section, but we elide this step for simplicity.

To implement neural resampling, we use PFNs~\cite{Komiske:2018cqr}, where each event is represented as a variable-length set of four-vectors that correspond to the muons, neutrinos, and clustered jets.
The \textsc{PdgID}~\cite{PhysRevD.98.030001} of the muons and neutrinos are used as a per-particle feature in the PFN, while the $b$-jets (ISR jets) are given the per-particle feature of 1 (0).
In principle, we could have applied neural resampling directly to the final-state hadrons, but since we are only going to plot jet-related quantities, it makes sense to perform jet clustering before neural resampling.

Note that the phase space that is being reweighted here is variable dimensional, with at least 20 dimensions from the 3-momenta for 6 particles and 2 jet masses.
These data are constrained to an 18-dimensional manifold after considering transverse momentum conservation and are approximately constrained to a 14-dimensional manifold after accounting for mass-shell conditions.
Many events have far more dimensions due to additional jets.
Thus, this is a highly non-trivial test of whether neural resampling can yield sensible results across a multi-dimensional and variable-dimensional phase space.

In \Fig{matched_isr_dist}, we plot the distribution of transverse momentum ($p_T$) of the leading ISR jet.
As expected, neural resampling has matched the bulk of this distribution up to statistical uncertainties.
In the tails, where training data are more sparse, there are larger variations, but we emphasize that these results were obtained ``out of the box'' with no attempt to optimize training parameters.

The distribution of event weights is shown in \Fig{matched_isr_weight}.
The original sample had both positive and negative weights, but neural resampling has yielded a positive weight distribution as desired.
With the optimal choice of $K$, the event weights extend out to about 10, which reflects the initial inefficiency of event generation with positive and negative weights.

The uncertainties are shown in \Fig{matched_isr_error}, again using the standard Monte Carlo estimate from \Eq{MCuncertainty}.
With $K = 1$, the uncertainties are underestimated throughout the phase space, but with the optimal value of $K$, they are captured correctly.
Correspondingly, the optimal value of $K$ yields a substantially lower number of events, as shown in \Fig{matched_isr_events}, indicating a large gain in downstream computational efficiency.
In particular, from the original 2M events, only about 600k remain after neural resampling, with no change to the statistical properties.

To highlight that neural resampling yields sensible results across the whole (unbinned) phase space, we plot the distribution of the number of ISR jets with $p_T$~$>$~10~GeV in \Fig{matched_njet}.
We emphasize that no retraining is needed here, since these plots are based on the same weights already shown in \Fig{matched_isr_weight}.
Obtaining results like this that work for any observable of interest would be challenging with a binned approach.
This case study therefore suggests that neural reweighting will be a powerful tool for efficient use of Monte Carlo generators at colliders.

\section{Conclusions and Outlook}
\label{sec:conclusions}

This paper has introduced neural resampling, a neural-network-based extension of the binned positive resampler proposed in \Ref{andersen2020positive}.
By exploiting the ability of neural networks to approximate likelihood ratios, our new approach is able to eliminate negative weights without binning and with access to potentially high-dimensional (and variable-sized) phase spaces.
Furthermore, neural resampling preserves statistical uncertainties with minimal overhead and no adjustable parameters.
This uncertainty preservation scheme is general and can also be applied to binned (non-neural network) resampling.

Given the growing availability of higher-order corrections and the computational demands of detector simulations, there is a need to make the best use of limited resources.
The neural resampler is able to preserve the statistical power of these precision calculations while decreasing downstream computational demands.
For the future, it would be interesting to study whether neural resampling could be incorporated more directly into the Monte Carlo generation process.
This could lead to further computational gains, particularly if used in concert with emerging neural-network-based phase space integration techniques.

Beyond just computational efficiency, there may be other advantages of neural resampling.
Many analysis strategies that consider the full unbinned log likelihood over events require per event weights to be positive~\cite{Nachman:2019dol}, which is guaranteed by neural resampling (assuming non-negative cross section).
Monte Carlo generators increasingly have the ability to keep track of uncertainties via weight variations~\cite{Mrenna:2016sih,Bothmann:2016nao}, which may be straightforward to incorporate into neural resampling via parametrized networks~\cite{Baldi:2016fzo,Andreassen:2019nnm}.
Finally, neural resampling ensures that the event weights are a true function of phase space and not multivalued maps, even allowing the weight function to be differentiated.
This feature of the learned weight function may turn out to be beneficial for other aspects of data processing and analysis.

\section*{Code and Data}

The code for this paper can be found at \url{https://github.com/bnachman/NeuralPositiveResampler}.  The top quark datasets are available upon request.

\section*{Acknowledgments}

We would like to thank A.~Andreassen, P.~Komiske, and E.~Metodiev for many helpful discussions about reweighting with neural networks.
We thank J.~Andersen, C.~Gutschow, A.~Maier, and S.~Prestel for helpful conversations about \Ref{andersen2020positive}.
BN was supported by the U.S.~Department of Energy (DOE) Office of Science under contract DE-AC02-05CH11231.
JT was supported by the U.S.~DOE Office of High Energy Physics under contract DE-SC0012567.
BN would also like to thank NVIDIA for providing Volta GPUs for neural network training.

\bibliography{HEPML,myrefs}

\end{document}